\documentclass[aps,pra,twocolumn,groupedaddress,showpacs]{revtex4}

\usepackage{graphicx,soul}
\usepackage{subfigure}
\usepackage{hyperref}
\usepackage[latin1]{inputenc}
\usepackage{amsfonts,amsmath,amssymb}
\usepackage{appendix}
\usepackage{color}
\usepackage{epstopdf}

\newcommand{\QPH}[1]%
{\href{http://arxiv.org/abs/quant-ph/#1}{quant-ph/{#1}}}
\newcommand{\mean}[1]{\left\langle#1\right\rangle}

\newcommand{\red}[1]{\textcolor{black}{#1}}

\newcommand{\ket}[1]{\left|#1\right>}

\usepackage{layout}

\begin{document}

\title{Double-lambda microscopic model for entangled light generation by four-wave-mixing}

\author{Q. Glorieux}
\email{quentin.glorieux@univ-paris-diderot.fr}
 \affiliation{Laboratoire
 Mat\'{e}riaux et Ph\'{e}nom\`{e}nes Quantiques,
  UMR 7162, Universit\'{e} Paris-Diderot CNRS\\
 10, rue A. Domon et L. Duquet, 75013 Paris, France}
 
\author{R. Dubessy}
 \affiliation{Laboratoire
 Mat\'{e}riaux et Ph\'{e}nom\`{e}nes Quantiques,
  UMR 7162, Universit\'{e} Paris-Diderot CNRS\\
 10, rue A. Domon et L. Duquet, 75013 Paris, France}

\author{S. Guibal, L. Guidoni, J.-P. Likforman}

 \affiliation{Laboratoire
 Mat\'{e}riaux et Ph\'{e}nom\`{e}nes Quantiques,
  UMR 7162, Universit\'{e} Paris-Diderot CNRS\\
 10, rue A. Domon et L. Duquet, 75013 Paris, France}
 
 \author{T. Coudreau}
  \affiliation{Laboratoire
  Mat\'{e}riaux et Ph\'{e}nom\`{e}nes Quantiques,
   UMR 7162, Universit\'{e} Paris-Diderot CNRS\\
  10, rue A. Domon et L. Duquet, 75013 Paris, France}

\author{E. Arimondo}

\affiliation{INO-CNR, Dipartimento di Fisica {\it E. Fermi},
Universit\`a di Pisa, Largo Pontecorvo 3, I-56127 Pisa, Italy}

\date{\today}

\begin{abstract}

Motivated by recent experiments, we study four-wave-mixing in an atomic double--$\Lambda$ system driven by a far-detuned pump.
Using the Heisenberg-Langevin formalism, and based on the microscopic properties of the medium,  we calculate the classical and quantum properties of seed and conjugate beams beyond the linear amplifier approximation.
A continuous variable approach gives us access to relative-intensity noise spectra that can be directly compared to experiments.
Restricting ourselves to the cold-atom regime, we predict the generation of quantum-correlated beams with a relative-intensity noise spectrum well below the standard quantum limit (down to -6 dB).
Moreover entanglement between seed and conjugate beams measured by an inseparability down to 0.25 is expected. 
This work opens the way to the generation of entangled beams by four-wave mixing in a cold atomic sample.
\end{abstract}

\maketitle%

\section{Introduction}
Four--wave mixing (FWM) has been identified early on as a very efficient process to generate intense, non classical beams \cite{Yuen79}.
Indeed, the first \red{experimental} demonstration of squeezed light was made using FWM in a sodium atomic beam almost 25 years ago \cite{Slusher1985}.
More recently, FWM in three--level $\Lambda$ systems was predicted to generate squeezing by using a counter--propagating geometry ~\cite{Shahriar1998,Lukin1998,Lukin1999,Lukin2000}.
In addition Electromagnetically Induced Transparency (EIT) in  double-$\Lambda$ atomic systems adds flexibility to the control of the FWM process, leading to large parametric gain and oscillations~\cite{Zibrov99,Lukin01}. 
\red{With such a system}, van der Wal {\it et al.} generated intensity correlations between light pulses with a corresponding 0.2~dB of \red{noise reduction below the standard quantum limit (SQL)}~\cite{vanderWal2003}.
FWM in a double-$\Lambda$-system in a hot rubidium vapour and with a co--propagating laser geometry was implemented in refs.~\cite{McCormick2007,McCormick2007b,McCormick2008,Pooser2008, Boyer2008,Marino2009} and by some of us in ref.~\cite{Glorieux2010} to generate bright correlated beams with a high degree of intensity correlations. 
\\
\indent 
In order to produce paired photons / twin beams tuned near an atomic resonance, efficient FWM in atomic samples with a low parametric gain requires large optical depths.
On the other side cold atom samples provide a precise control of atomic parameters as inhomogeneities and decoherence processes.
Thus, the  combination of double-$\Lambda$ EIT and cold atoms  has allowed the observation of FWM at ultra-low optical powers.
An optically thick cesium MOT allowed Kuzmich {\it et al.}~\cite{Kuzmich2003} to generate nonclassical photon pairs with a programmable delay. Generation and control of narrow-band paired photons were demonstrated in a rubidium MOT~\cite{Braje2004,Balic2005,Kolchin2006,Kolchin2008} by reaching a high optical depth (OD), around 60~\cite{Du2008}.
Correlated photons were \red{also} produced by using cold atoms inside a cavity~\cite{Thompson2006} .\\
\indent \red{Following \cite{Lukin}}, recent publications~\cite{Kolchin2007,Ooi2007,DuRubin2008} have theoretically explored the time correlations between photons observed in cold atom experiments.
Propagation of the generated photons through the optically thick medium and Langevin forces represent the key elements of these theoretical analyses. 
Following the seminal papers \cite{Lukin1998,Lukin1999,Lukin} and the analysis of  \cite{Kolchin2007}, we investigate the generation of quantum correlated beams \red{in co--propagating beam geometry}.
This work presents a detailed theoretical treatment of FWM in an optically thick medium composed by cold atoms described by a four level model,  in a double-$\Lambda$ configuration.

\indent The main target is to explore the production of correlated beams in the atom-laser configuration experimentally investigated in refs.~\cite{McCormick2007,McCormick2007b,McCormick2008,Pooser2008,Boyer2008,Marino2009,Glorieux2010}, but using a cold atom sample as in refs.~\cite{Balic2005,Kolchin2006,Kolchin2008,Du2008}. 
While the theoretical descriptions ~\cite{Kolchin2007,Ooi2007,DuRubin2008}, concentrated on the few photon regime, the experiments discussed in refs.~\cite{McCormick2007,McCormick2007b,McCormick2008,Pooser2008,Boyer2008,Marino2009,Glorieux2010} require to tackle the large photon number regime where a description based on continuous variables is required \cite{Bachor2004,Braunstein2005}.
Within this approach, we derive the quantum noise frequency spectra and quantify the quantum entanglement between the generated beams.\\
\indent Several additional original points of our analysis should be listed. 
First, we investigate the pump-seed co--propagating geometry, a setup which enabled to produce large intensity quantum correlations in the experiments ~\cite{McCormick2007,McCormick2007b,McCormick2008,Pooser2008,Boyer2008,Marino2009,Glorieux2010}.
Furthermore, in contrast to the phenomenological description of ref.~\cite{McCormick2008}, gain and propagation losses are intrinsically included in our microscopic approach. 
As the most important result, our analytical and numerical analysis  predicts that a very large degree of intensity quantum correlations can be achieved also in a cold atom FWM experiments, comparable to that achieved in~refs \cite{McCormick2007,McCormick2007b,Glorieux2010}.
We also investigate the ability of such a medium to simultaneously generate large phase anticorrelations, leading to entanglement, a key ressource for quantum information~\cite{Braunstein2005}.

\indent The paper is organized as follows. Section \ref{sec:micr-model-heis} introduces the microscopic model as described by Heisenberg-Langevin equations.
Section \ref{sec:analytical-solution} discusses the analytical solution for the propagating quantum fields.
In Section \ref{sec:meas-quant} we derive the mean values of the generated beams. 
We introduce and calculate in Section~\ref{sec:results} the intensity quantum correlations and phase quantum anti--correlations spectra between seed and conjugate beams including the Langevin forces contribution.
Finally, in Section \ref{sec:discussion}, we discuss the effect of decoherence processes and \red {compare the results to hot vapor experiments.}
\section{Microscopic model \textbf{--} Heisenberg-Langevin equations}
\label{sec:micr-model-heis}

\begin{figure}[ht]
\centerline{\scalebox{0.5}{\includegraphics{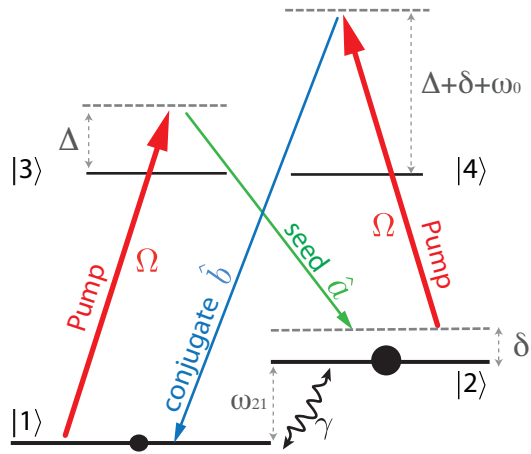}}}
\caption{(color online) Level diagram with relevant detunings.
The two large red arrows indicate the pump coupling between levels $\ket{1}$ and $\ket{3}$ (resp. $\ket{2}$ and $\ket{4}$), with equal Rabi frequency $\Omega$. $\delta$ represents the two-photon detuning on both $\Lambda$ subsystems. The strong pump and weak seed configuration leads to optical pumping of the atoms into level $\ket{2}$. \label{levels}}
\end{figure}

We consider a collection of  atoms  having the double $\Lambda$ type 4--level scheme of  Fig.~\ref{levels}. 
The four atomic levels $\ket{u}$,  are defined by their energies $E_u$, and we introduce the angular frequencies  $\omega_{uv}=(E_u-E_v)/\hbar$,  $u,v \in\{1,2,3,4\}, u>v$. 
The populations of levels $\ket{3}$ and $\ket{4}$ decay with a rate $\Gamma$.
The atomic coherence between levels $\ket{1}$ and $\ket{2}$ decays with rate $\gamma$ and we neglect $\ket{1}$ and $\ket{2}$ population decays since we consider cold atomic samples.

\indent The atoms interact with three optical fields, pump (angular frequency $\omega_p$) and seed (angular frequency $\omega_a$) driving the first $\Lambda$ subsystem ($|1\rangle \to \ket{3} \to  \ket{2}$), conjugate (angular frequency $\omega_b$) and pump driving the second $\Lambda$ subsystem  ($|1\rangle \to \ket{4} \to  \ket{2}$).  
The detuning of the pump field from the $\ket{1} \to \ket{3}$ transition is denoted~$\Delta$.
The pump and the seed are near-resonant with the $\ket{1} \to \ket{3} \to \ket{2}$  Raman transition with a two--photon detuning $\delta$. 
The double-$\Lambda$ is closed by the conjugate and pump beams driving the $\ket{1} \to \ket{4} \to \ket{2}$  Raman transition, assuming energy matching.
As shown in Fig.\ref{levels} the pump laser is detuned by $\Delta +\omega_0+\delta$ from the $\ket{2}  \to \ket{4}$ transition, where $\omega_0=\omega_{21}-\omega_{43}$. 

\indent The interactions of the pump beam with the atoms are described semi-classically. 
For the sake of simplicity we limit our study to a 1D model with propagation along the $z$-axis.
The seed and conjugate beams are described by two slowly varying quantum--mechanical operators;  the annihilation (resp. creation) operators being denoted $\hat a$ (resp. $\hat a^\dagger$) and $\hat b$ (resp. $\hat b^\dagger$).

Following the approach  of ref.~\cite{Lukin}, the properties of the medium are described by collective, slowly--varying operators   $\hat \sigma_{uv}(z,t)$ averaged over small layers denoted by their position $z$. 
For each layer containing a large number of atoms $N_z$,  we define the following population/coherence operators $\hat{\sigma}_{uv}$: 
\begin{equation}
  \label{eq:2}
  \hat \sigma_{uv}(z,t) = \frac{1}{N_z} \sum_{j=1}^{N_z} |u_j  \rangle \langle v_j | e^{(- i \tilde \omega_{uv} t+i k_{uv} z)},
\end{equation} where $\tilde \omega_{31} = \tilde \omega_{42}= \omega_p $, $\tilde \omega_{32}=\omega_a$, $\tilde \omega_{41}=\omega_b$, $\tilde \omega_{21}=~\tilde \omega_{43}=~\omega_{p}~-~\omega_{a}$ . 
The $k_{uv}$ are the projections of the wavevectors $\vec k_{uv}$ on the $z$ axis with $\vec k_{31}$ and $\vec k_{42}$ equal to the pump wavevector, $\vec k_{32}$ equal to the seed wavevector, and $\vec k_{41}$ equal to the conjugate wavevector, $\vec k_{21} =  \vec k_{23} - \vec k_{13}$ and $\vec k_{43}=\vec k_{42} - \vec k_{32}$, with the convention $\vec k_{uv}=-\vec k_{vu}$.

\indent We write  the laser--atom interaction Hamiltonian $\hat{V}$ in the rotating wave approximation, omitting for simplicity the $z$ and $t$ dependencies of the atomic operators:

\begin{eqnarray}
\hat{V}&&=-\frac{\hbar N}{L}\int^L_0 dz [(\omega_0+\Delta+\delta)\ \hat \sigma_{44}+\Delta \hat \sigma_{33}+\delta \hat \sigma_{22}\\
\nonumber&&+g_a\hat{a}(z,t) \hat  \sigma_{32}  +g_b \hat{b}(z,t) \hat  \sigma_{41}+\frac{\Omega}{2}(\hat \sigma_{31}+\hat \sigma_{42})]+H.c.,
\end{eqnarray}
where $N$ is the number of atoms in the quantization volume $V$ of length  $L$. 
$g_{a,b}$ are the atom -- field coupling constants:  $g_a=\frac{\wp_{32}\varepsilon_{a}}{\hbar}$ and  $g_b=\frac{\wp_{41}\varepsilon_{b}}{\hbar}$ where $\wp_{uv}$ is the transition dipole moment and $\varepsilon_{a,b}=\sqrt{\frac{\hbar\ \omega_{a,b}}{2 \epsilon_0 V }}$  is the electric field of a single photon.
\red{In the following, we consider $g_a=g_b=g$}.
Finally, $\Omega=\frac{2\wp E}{\hbar}$ is the pump laser Rabi frequency with $E$ the pump laser electric field amplitude.
We assume the transition dipole moment  $\wp$  to be equal for the two pump-driven transitions $|1\rangle \rightarrow|3\rangle$ and $|2\rangle \rightarrow|4\rangle$. 

\red{As described in} ref.~\cite{Kolchin2007}, the evolution of atomic operators is determined by the following Heisenberg-Langevin equations:
 \begin{equation}
\left(\frac{\partial}{\partial t}+\gamma_{uv}\right) \hat \sigma_{uv}=\frac{i}\hbar [\hat{V}, \hat \sigma_{uv}]+  \hat  r_{uv}+\hat F_{uv},
\end{equation}
where $\gamma_{uv}$ are the atomic dephasing rates and $\hat r_{uv}$ are the source terms produced by spontaneous emission as defined in ~\cite{Kolchin2007}.
 Following the ref.~\cite{Ooi2007}, the Langevin operators $\hat F_{uv}$  are characterized by
   \begin{equation}
  \mean{\hat F_{uv}(z,t)}=0,
  \end{equation}
  and 
  \begin{equation}
  \mean{\hat F_{uv}^\dag(z,t)\hat{F}_{u'v'}(z',t')}=2D_{uv,u'v'}\delta(t-t')\delta(z-z').
 \label{diffusion}
  \end{equation}
Equation ~\eqref{diffusion} defines the 256 diffusion coefficients $D_{uv,u^\prime v^\prime}$;  we show in Appendix \ref{app:3} that only a few of them will be required in next section.

\indent Following  the  approach of  refs.~\cite{Lukin1999,Zibrov99,Kolchin2007}, the description of the atom--laser system is completed by a set of nonlinear, coupled differential equations describing the propagation and temporal evolution of the quantum field operators
\begin{eqnarray}\label{eq:propa1}
\left(\frac{\partial}{\partial t}+c\frac{\partial}{\partial z}\right) \hat a(z,t)&=&i g \mathcal{N} \hat\sigma_{23}(z,t),\\
  \left(\frac{\partial}{\partial t}+c\frac{\partial}{\partial z}\right) \hat{b}^\dag(z,t)&=& - i g \mathcal{N} \hat\sigma_{41}(z,t)\label{eq:propa2} ,
\end{eqnarray}
where $\mathcal{N}=N/V$ is the atomic density.

\section{Analytical solution}
\label{sec:analytical-solution}
The system evolution  is described  by a set of non--linear, coupled differential equations for quantum fields and atomic operators which require further approximations to be solvable analytically.
We assume a pump beam highly saturating the medium and very intense compared to the seed and conjugate beams.
Within these assumptions the  populations $ \hat{\sigma}_{11},\,   \hat{\sigma}_{22},\,  \hat{\sigma}_{33},\, \hat{\sigma}_{44}$  and the coherences $ \hat{\sigma}_{31},\, \hat{\sigma}_{13},\,  \hat{\sigma}_{42},\,  \hat{\sigma}_{24}$ are dominantly driven by the pump beam.
  In this case the complete non--linear differential equations system can be separated into two subsystems. 
In the first subsystem \red{(Eq. A.1)}, the effect of seed and conjugate can be neglected and \red{these equations are} solved in the steady state. 
The corresponding solutions are given in Appendix \ref{app:2}.
 These solutions are then injected in \red{Eq.(B1)} which determine the coherences $  \hat{\sigma}_{14},\, \hat{\sigma}_{41},\,  \hat{\sigma}_{23},\,  \hat{\sigma}_{32},\hat{\sigma}_{12},\, \hat{\sigma}_{21},\,  \hat{\sigma}_{34},\,  \hat{\sigma}_{43}$ as a function of $\hat{a}$ and $\hat{b}^\dag$ as described in detail in Appendix \ref{app:3}.\\
 \indent Eqs.~(\ref{eq:propa1}) and \eqref{eq:propa2} can then be written in a closed form and solved analytically in the Fourier space. 
The Fourier transforms of $\hat{a}(t,z),\, \hat{b}^\dag(t,z)$ will be denoted    $\hat{a}(\omega,z),\, \hat{b}^\dag(\omega,z)$,  where for simplicity we have used the same notation for operators in the time and frequency domains.
Notice our notation for the Fourier transform of an operator and its conjugate 
\begin{equation}
\hat{a}(\omega)=\int_{-\infty}^\infty \hat{a}(t)\ e^{i\omega t}\ dt\ ;\ \hat{a}^\dag(\omega)=\int_{-\infty}^\infty \hat{a}^\dag(t)\ e^{i\omega t}\ dt,
\end{equation}
leading to  $[\hat{a}(\omega)]^\dag=\hat{a}^\dag(-\omega)$.
The Fourier transforms of the quantum operators satisfy then the following equations:
 \begin{eqnarray}
\frac{1} {k_{32}} \frac{\partial}{\partial z} \hat a(\omega)&=& \eta_{a}(\omega) \hat{a}(\omega)+\kappa_{a}(\omega) \hat{b}^\dagger(\omega) + \hat{\mathcal{F}}_a^{at.} (\omega),\label{eq9}\\
\frac{1} {k_{41}}  \frac{\partial}{\partial z} \hat{b}^\dag(\omega)&=&\kappa_{b}(\omega)\hat{a}(\omega)+ \eta_{b}(\omega)\hat{b}^\dagger(\omega)+ \hat{\mathcal{F}}_{b^\dagger}^{at.} (\omega),\hspace{0.5cm}\label{eq10}
\end{eqnarray}
where $\eta_{a}(\omega)$ (resp. $\eta_{b}(\omega)$) is the complex refractive index of the medium "dressed" by the pump laser for the mode $\hat{a}$ (resp. $\hat{b}^\dagger$). $\kappa_{a}(\omega)$ (resp. $\kappa_{b}(\omega)$) is the parametric conversion coefficient from mode $\hat{a}$ to $\hat{b}^\dagger$ (resp. $\hat{b}^\dagger$ to $\hat{a}$), $\hat{\mathcal{F}}_{a}^{at.}$ and $\hat{\mathcal{F}}_{b^\dagger}^{at.}$ are the Langevin terms originating from the atomic Langevin forces and can be calculated from Eq. \eqref{flucuat}.

\indent  Following a standard approach of quantum optics~\cite{Davidovich1996}, we use the input/output formalism to compute the quantum operators of seed and conjugate fields.
We introduce the input operators, $\hat{a}_{in}$ and $\hat{b}^{\dag}_{in}$ for 
 $z=0$ and the output operators $\hat{a}_{out}$ and $\hat{b}^{\dag}_{out}$ for $z=L$. 
 The formal solution of the propagation equations \eqref{eq9}-\eqref{eq10} is given by
\begin{equation}
\begin{bmatrix}\hat{a}_{out}(\omega) \\\hat{b}_{out}^\dag(\omega)\end{bmatrix}
=\begin{bmatrix}A(\omega) & B(\omega) \\C(\omega) & D(\omega)\end{bmatrix}
\left(\begin{bmatrix}\hat{a}_{in}(\omega) \\\hat{b}_{in}^\dag(\omega)\end{bmatrix}
+\begin{bmatrix}\hat{F}_{a,out}(\omega)\\\hat{F}_{b,out}(\omega)\end{bmatrix}\right)\label{eq:8}
\end{equation}
where the coefficients $A(\omega), B(\omega),C(\omega),D(\omega)$ and the Langevin terms $\hat{F}_{b,out}(\omega),\hat{F}_{a,out}(\omega)$ are defined in Appendix \ref{app:C}.
 Eq.\eqref{eq:8} allows us to compute both the output fields mean values ($\mean{\hat a_{out} }, \langle{\hat b_{out}^\dag }\rangle$) and their fluctuations ($\delta{\hat a_{out} }, \delta{\hat b_{out}^\dag })$.

 As in typical FWM experiments, in the following we will suppose that the input on the conjugate mode $\hat{b}$ is the vacuum ($\langle{\hat{b}_{in}}\rangle=0$) and that the input on the seed mode is an arbitrary field of mean value $\mean{\hat{a}_{in}}$. The average values of the output seed and conjugate fields are given by
\begin{eqnarray}
\mean{\hat{a}_{out}(\omega=0)}&=&A(0)\mean{\hat{a}_{in}}\\
\mean{\hat{b}_{out}^\dagger(\omega=0)}&=&C(0)\mean{\hat{a}_{in}}. 
\end{eqnarray}
 \red{We define then $G_a=|A(0)|^2$ as the seed gain and $G_b=|C(0)|^2$ as} the ratio between the conjugate output and seed input intensities.

\section{Classical properties of the medium}
\label{sec:meas-quant}


\begin{figure}[ht]
\centerline{\scalebox{0.65}{\includegraphics{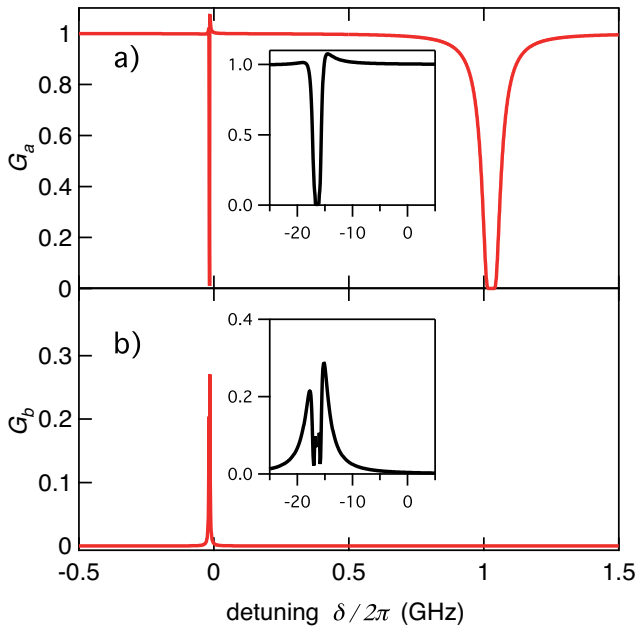}}}
\caption{Spectra of the $G_a$ seed beam gain in a) and of the conjugate beam gain $G_b$ in b) as a function of the two photon detuning $\delta$. The parameters used for the plot are : $\gamma/2\pi=10$~kHz; $\Omega/2\pi=0.3$~GHz;
 $\Delta/2\pi=1$~GHz.
The seed beam dip for $\delta\simeq\Delta$ in a) is originated  by absorption  on the $|2\rangle \to |3\rangle$ transition. 
 Expanded views of $G_a$ and $G_b$ gains around $\delta=0$ are shown in the insets.
  The peculiar line-shapes are the result of a nontrivial interplay between the two processes discussed in the text.}
\label{gain}
\end{figure}

\red{The mean value expressions} for the seed and conjugate output operators allow us to calculate the output intensities which can be experimentally measured. 
The parameters used in the following for laser intensities and detunings are derived from hot vapors experiments such as~\cite{McCormick2007, McCormick2007b,McCormick2008,Pooser2008,Boyer2008,Marino2009,Glorieux2010}, \red{so that} in our numerical  calculations, we will use   $0.7<\Delta /2\pi  <3$~GHz and $0.3<\Omega/2\pi<2$~GHz.
\indent We consider the $D_1$ transition of $^{85}$Rb: lasers wavelength 795~nm; $\omega_{0}/2\pi=3$~GHz; natural linewidth $\Gamma/2\pi= 5.7$~MHz; resonant absorption cross section $\sigma_0=10^{-9}$~cm$^2$ \cite{Steck2009}.
For the optical depth defined by $OD=\mathcal{N}\sigma_0 L $, we use the value of 150 reachable by improving a cold-atom configuration such that of ref.~\cite{Du2008}.
In cold atom experiments, magnetic field inhomogeneity and atomic collisions may determine the $\gamma$ decoherence rate.
In a state-of-the-art experiment, $\gamma$ depends mainly on the relative phase stability of the lasers and is as low as $\gamma / 2\pi\simeq 10$~Hz~\cite{Bloch}.
However, as specifically discussed in Sec.~\ref{sec:dec-rate}, our \red{results} shows a weak dependence on this parameter and  $\gamma / 2 \pi=10$~kHz will be usually assumed.\\
\indent  Fig.~\ref{gain} reports numerical simulation results for the seed and conjugate gains as function of the two-photon detuning $\delta$. 
 As described in ref.~\cite{Lukin2000} different elementary processes contribute to the system dynamics.
Effective coherent photon redistribution from pump to seed and conjugate occurs when the four-photon resonance condition is fulfilled ($\delta\simeq 0$). 
 Let us note that the same resonance condition holds for a Raman absorption process involving the seed beam. 
While it is quite obvious that mode $\hat{b}$ will always be amplified as soon as the non--linearity exceeds the absorption loss, the behaviour of mode $\hat{a}$ is more complicated and relies on the interference between redistribution and Raman processes.
The spectrum profile of the seed gain (Fig.~\ref{gain}a) is characterized by  absorption  at  $\delta \simeq \Delta$, with a width imposed by $\Gamma$ and the propagation through the optically thick medium.
For $\delta\simeq 0$, the seed gain spectrum displays a sharp profile containing both gain and absorption contributions (Fig. \ref{gain}a inset).
The seed absorption is due to the Raman process on the transition $|2\rangle\rightarrow |1\rangle$ involving one seed photon and one pump photon, \red{while} the gain can be attributed to the FWM process. 
For a pump laser in resonance with one ''arm'' of the $\Lambda$ transition the EIT profile assumes the characteristic lineshape of a narrow dip inside a Lorentzian profile. 
In an EIT configuration with  the pump laser detuned from the resonance, the absorption profile of the seed field becomes asymmetric about the two-photon resonance assuming a characteristic Fano-like profile as in the experiments of refs.~\cite{Alzetta1979,Kaivola1985,Zhu1996,Stalgies1998} and the theoretical analysis of ref.~\cite{Lounis1992}.
The $\delta\simeq 0$  profile of $G_a$ is similar to the EIT Fano profiles reported in the above references, except that $G_a \geq 1$ in a narrow frequency range.
The precise positions of the seed absorption and Fano-like feature are given by the AC Stark shifts induced by the pump  laser.

\indent The $G_b$ spectrum,  shown in Fig. \ref{gain}b reports a reduction of the gain in the region $\delta\simeq 0$ with a width matching that of the dip in the $G_a$ spectrum.
\red{In} a naive description, conjugate gain \red{does not increase along} the propagation when the seed is completely depleted.

\section{Quantum properties of the system}\label{sec:results}

\subsection{Noise spectra}
Now we focus our attention to the quantum properties of the system.
We follow the standard approach of the continuous variable formalism  \cite{Davidovich1996} to calculate the noise spectra.
For the seed mode $\hat{a}$ we introduce the following amplitude and phase quadrature fluctuations components,  $ \delta \hat x_{a}$ and  $\delta \hat p_{a}$ respectively:
\begin{eqnarray}
  \label{eq:1}
  \delta \hat x_{a} &=& \delta \hat{a} e^{-i \varphi_a}+ \delta \hat{a}^\dagger e^{i \varphi_a}\\
  \label{eq:3}
  \delta \hat p_{a} &=& -i (\delta \hat{a} e^{-i \varphi_a}- \delta \hat{a}^\dagger e^{i \varphi_a})
\end{eqnarray}
where $\varphi_a$ is  the phase  of the mean field $\mean{\hat{a}}$.
We define accordingly the amplitude quadrature fluctuations for the conjugate mode $\hat{b}$, $  \delta \hat x_{b}$ and   $\delta \hat p_{b}$. 
Following the  Wiener--Khintchine theorem, the spectrum of a given quantity $\hat u$ can be evaluated \red{by} taking the Fourier transform of its autocorrelation function~\cite{Davidovich1996}:
\begin{equation}\label{wk}
S_{\hat u}(\omega) 2\pi \delta (\omega - \omega')= \mean{\hat u(\omega) \hat u^\dag (\omega')}
\end{equation}

\indent We analyze the  FWM  configuration of refs.~\cite{McCormick2007,McCormick2007b,McCormick2008,Pooser2008,Boyer2008,Marino2009,Glorieux2010} with a coherent state for the seed input of mean value amplitude $\alpha$. In agreement with the symmetric ordering of the operators \cite{Davidovich1996} to be used hereafter, the covariance matrix for the input seed field is given by
\begin{widetext}
\begin{eqnarray}\label{covmat}
\begin{bmatrix}
\mean{ \delta\hat x_{a,in}(\omega)  \delta\hat x_{a, in}^\dagger(\omega')} & \mean{ \delta\hat x_{a, in}(\omega)  \delta \hat p_{a, in}^\dagger (\omega')} \\
\mean{ \delta\hat p_{a, in}(\omega)  \delta\hat x_{a, in}^\dagger(\omega')} & \mean{ \delta\hat p_{a, in}(\omega)  \delta \hat p_{a, in}^\dagger (\omega')} 
\end{bmatrix} = 2\pi \delta(\omega+\omega')
\begin{bmatrix}
1 & 0 \\0 & 1
\end{bmatrix},
\end{eqnarray}
\end{widetext}
and a similar definition applies for covariance matrix of the conjugate mode but with a zero mean input power.
Notice that, according to \cite{Fabre90,Ekert91}, the choice of ordering of the operators should not influence the calculated spectra.

Using Eq. \eqref{eq:8} for the propagation  of the quantum operators for the seed and conjugate fields and the input covariance matrix defined in Eq. \eqref{covmat}, the quantum noise spectra for the quadrature amplitude of the fields are given by
\begin{widetext}
  \begin{eqnarray}\label{eq:sol_a}
  \nonumber  S_{\hat x_{a,out}}(\omega)  &=\frac{|\mean{\hat{a}_{out}}|^2}{2}& \left[ |A(\omega)|^2(1+D_{aa^\dag}(\omega))+|A(-\omega)|^2(1+D_{a^\dag a}(-\omega))\right.\\
      &&+ \left. |B(\omega)|^2(1+D_{b^\dag b}(\omega))+|B(-\omega)|^2(1+D_{bb^\dag}(-\omega))\right],\\
 \label{eq:sol_b} \nonumber  S_{\hat x_{b,out}}(\omega) &=\frac{|\mean{\hat{b}_{out}}|^2}{2}& \left[ |C(\omega)|^2(1+D_{aa^\dag}(\omega))+|C(-\omega)|^2(1+D_{a^\dag a}(-\omega))\right.\\
          &&+ \left. |D(\omega)|^2(1+D_{b^\dag b}(\omega))+|D(-\omega)|^2(1+D_{bb^\dag}(-\omega))\right],
  \end{eqnarray}
  \end{widetext}
where the \red{calculation method} of the four atom--driven Langevin diffusion terms $D_{ a a^\dag}$, $D_{a^\dag a}$ $D_{b^\dag b}$ and $D_{bb^\dag}$ is presented in Appendix \ref{appendix_langevin}. 
All those coefficients are, by construction, real and positive quantities and act as additional noise source terms on the quantum noise spectra. Their specific contribution will be discussed below. 


\indent The interest for the double-$\Lambda$ FWM system stems in particular from its ability to generate quantum field correlations. 
\red{Such correlations can concern a single quadrature of seed and conjugate fields (e.g. intensity correlations as in refs.~\cite{Heidmann87,Laurat05a}) or two conjugated quadratures (e.g. intensity correlations and phase anti-correlations).
In the second case, seed and conjugate fields can be entangled}
Let us note that while most FWM experiments investigated intensity squeezing, Ref.~\cite{Boyer2008} reports entanglement in a hot atomic vapour.
In order to \red{detect} the entanglement, we will here use the $\mathcal{I}(\omega)< 1$ sufficient criterion introduced in \cite{Duan2000,Simon2000} and based on 
 the inseparability parameter defined by 
 \begin{equation}\label{insep}
\mathcal{I}(\omega)=\frac 12 (S_{ x}^- + S_{p}^+).
\end{equation}
Here $S_{ x}^-$ is the intensity correlation spectrum and $S_{p}^+$ the phase anti--correlation spectrum both normalized to the standard quantum limit (SQL). 
$S_{ x}^-$ is derived by applying Eq. \eqref{wk} to 
\begin{equation}\hat u=\frac{1}{\sqrt 2}\left(\delta \hat x_{a,out}(\omega)-\delta \hat x_{b,out}(\omega)\right)\end{equation}
and normalizing it to \red{the sum of gain on mode $\hat{a} $ and mode $\hat b$:} $(G_a+G_b)$.
Correspondingly $S_{p}^+$ is obtained with  \begin{equation}\hat u=\frac{1}{\sqrt 2}\left(\delta \hat p_{a,out}(\omega)+\delta \hat p_{b,out}(\omega)\right) \end{equation} with the same normalization.
\red{Inseparability will be used below to give a higher bound for entanglement as discussed in ref.~\cite{Laurat05b}.
Eq.\eqref{insep} leads to the following expressions for the normalized intensity correlation spectrum and the normalized phase anti--correlation spectrum:}
\begin{widetext}\begin{footnotesize}
\begin{eqnarray}
\label{int_sqz}
S_{ x}^-(\omega)&=&
\frac{1}{2(G_a+G_b)} \left( 
(|A(0)^* A(\omega)-C(0)^* C(\omega)|^2)(1+D_{aa^\dag}(\omega))+(|A(0) A(-\omega)^*-C(0) C(-\omega)^*|^2)(1+D_{a^\dag a}(-\omega))\right.\nonumber \\
&+& \left.(|A(0)^* B(\omega)-C(0)^* D(\omega)|^2)(1+D_{b^\dag b}(\omega))+(|A(0) B(-\omega)^*-C(0) D(-\omega)^*|^2)(1+D_{bb^\dag}(-\omega))\right)\\
S_{p}^+(\omega)&=& \frac{1}{2(G_a+G_b)} \left(
(|A(0) C(\omega)-C(0)^* A(\omega)^*|^2)(1+D_{aa^\dag}(\omega))+(|A(0) C(-\omega)^*-C(0)^*A(-\omega)^*|^2)(1+D_{a^\dag a}(-\omega))\right.\nonumber\\\label{sol_phi}
&+&\left.(|A(0) D(\omega)-C(0)^* B(\omega)^*|^2)(1+D_{b^\dag b}(\omega))+(|A(0) D(-\omega)^*-C(0)^* B(-\omega)^*|^2)(1+D_{bb^\dag}(-\omega))\right)
\end{eqnarray}
\end{footnotesize}
 \end{widetext}
In contrast to the case of Eq. \eqref{eq:sol_a}, which is a sum of positive terms, a non-vanishing set of coefficients $A,B,C,D$ could exist in the Eq. $\eqref{int_sqz}$, such that the spectrum of the intensity difference $S_{ x}^-(\omega) \to 0$ \red{(same holds for Eq. 24)}.
For example, in the case of a system behaving as an infinite bandwidth ideal linear amplifier having gain equal to G ($|A(\omega)|^2=|D(\omega)|^2=G$ and $|B(\omega)|^2=|C(\omega)|^2=G-1$, and vanishing $D_{uv}$), the relative intensity noise spectra is given by \begin{equation}S_{ x}^-(\omega) = \frac{1}{2G-1},\end{equation} as reported in \cite{McCormick2008}.
Let us note that our model describes the frequency dependence of the noise spectra as well as the Langevin forces contribution through the coefficients $D_{ a a^\dag}$, $D_{a^\dag a}$ $D_{b^\dag b}$ and $D_{bb^\dag}$.

\begin{figure}[]
\scalebox{0.38}{\includegraphics{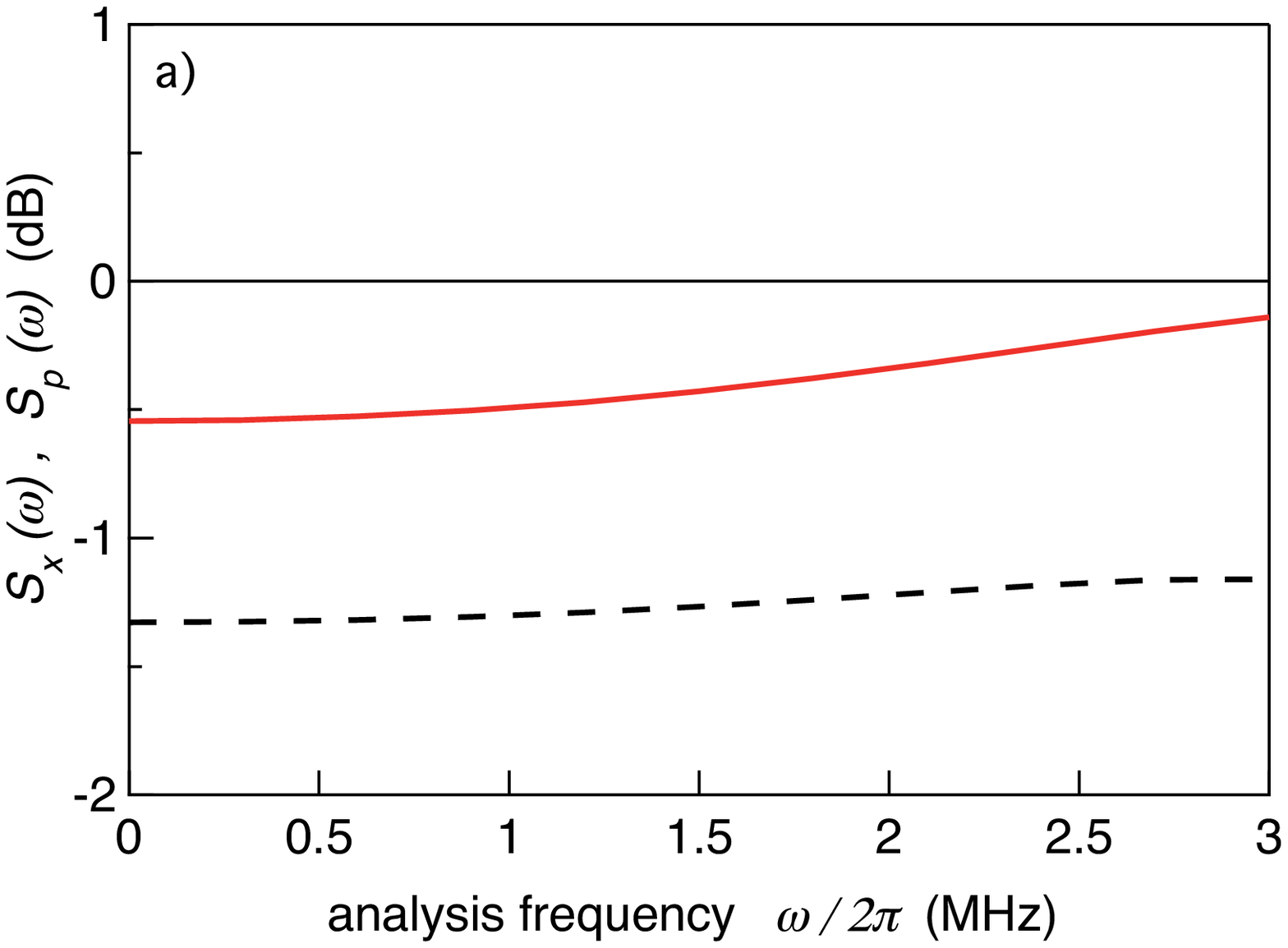}}\\
\scalebox{0.38}{\includegraphics{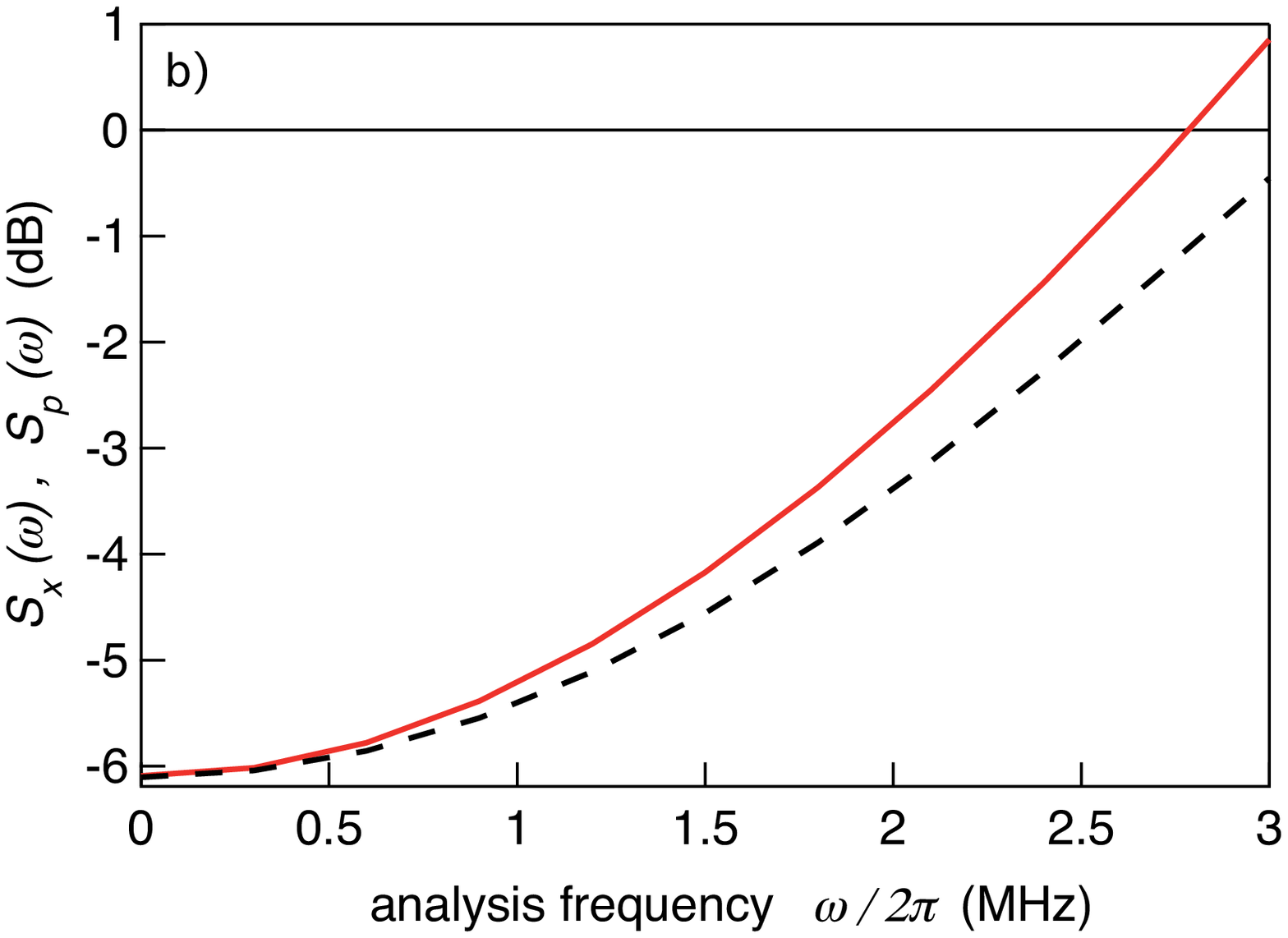}}
  \caption{(color online) Intensity correlations (black dashed) and phase anti--correlations (red continuous) spectra (in~dB) as a function of analysis frequency for a) $\Omega/2\pi=\Delta/2\pi=0.3$~GHz  and $\delta/2\pi=$-48 MHz and b) $\Omega/2\pi=\Delta/2\pi=2$~GHz and $\delta/2\pi=$-217 Mhz.
  Inseparability \red{can be obtained} by taking the half sum of the two correlation spectra. }
  \label{fig:squeezing}
\end{figure}

\indent Fig. \ref{fig:squeezing} displays the intensity correlations and phase anti--correlations spectra versus $\omega/2\pi$ (analysis frequency) for two values of $\Omega$ and $\Delta$.
For each set of values the two--photon detuning $\delta$ is chosen in order to optimize the noise reduction, close to the maximum gain value.
In both cases the output beams are entangled since they display both non classical intensity correlations and phase anti-correlations.
Our calculation shows that, within the explored range of parameters, at fixed $\Omega/\Delta$ the optimum entanglement increases with $\Omega$.
Let us note that both seed and conjugate output beams display an intensity noise above the standard quantum limit as reported in hot atomic vapor experiments \cite{McCormick2007}.

\subsection{Contribution of Langevin forces}

\begin{figure}[ht]
\scalebox{0.38}{\includegraphics{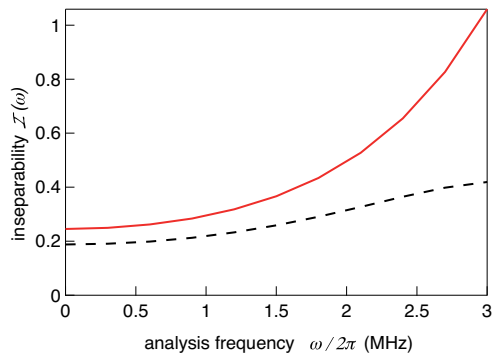}}\\
  \caption{(color online) Inseparabilty with (red line) and without (black dashed) the Langevin noise correction. Parameters as in Fig.~3b).}
  \label{fig:squeezingLangevin}
\end{figure}

Eqs. \eqref{int_sqz} and \eqref{sol_phi} show that the contribution of the Langevin forces is purely detrimental for entanglement. 
In a previous paper studying similar systems~\cite{Kolchin2007}, the contribution of Langevin forces was negligible.
In our system we found that it was not the case and we have specifically investigated their role.
Fig. \ref{fig:squeezingLangevin} shows the inseparability spectra versus the analysis frequency in presence and absence of the Langevin terms.
Their effect is small but clearly non--negligible and increases with $\omega$. Let us point out that, when neglecting the Langevin correction terms, our calculation leads to non--physical spectra for the mode $\hat{a}$, i.e., the noise can fall simultaneously below the SQL for both phase and intensity quadratures.
On the contrary, by including the Langevin terms, we always obtain, as mentioned above, excess noise on both quadratures.

\section{Discussion}
\label{sec:discussion}

Below, we discuss briefly the role of the ground state decoherence rate as well as \red{the comparison of the model to the results of a hot vapor experiment.}

\subsection{Role of the ground state decoherence rate}
\label{sec:dec-rate}
\begin{figure}[h]
\centerline{\scalebox{0.4}{\includegraphics{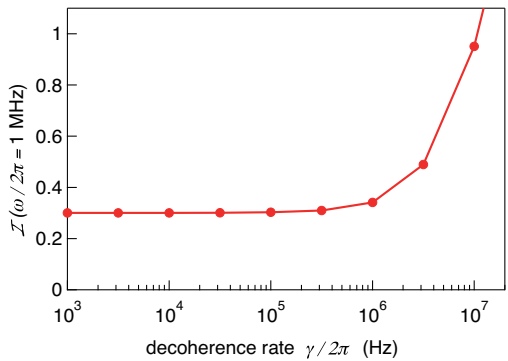}}}
  \caption{ Effect of the decoherence between the two ground state on the $\mathcal{I(\omega)}$ inseparability  at  $\omega/2\pi=1$ MHz. One can see that the process is robust to decoherence up to 1 MHz.  Parameters: $\Omega/2\pi=\Delta/2\pi= 2$ GHz, $\delta/2\pi=-217$ MHz.}
  \label{fig:gamma}
\end{figure}

Quantum correlations are notably sensitive to various decoherence mechanisms.
In Fig.~\ref{fig:gamma} we present the evolution of the maximal quantum correlations (in terms of the  $\mathcal{I}$ inseparability) as a function of the $\gamma$ decohernece rate.
While high relaxation rates ($\gamma/2\pi > 10$~MHz) do not allow for the observation of entanglement, the smooth behavior in the low $\gamma$ region  shows that the system is quite robust against mild decoherence mechanisms.
These observations show that coherences between hyperfine levels play a key role for the production of entanglement in FWM experiments.

\subsection{Comparison to hot atomic vapor experiments}

\begin{figure}[h]
\centerline{\scalebox{0.6}{\includegraphics{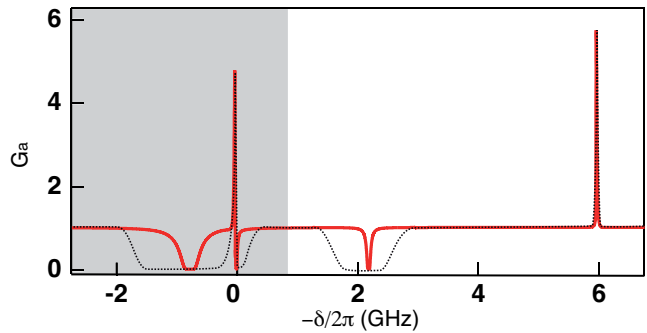}}}
  \caption{Gain spectrum $G_a$ of a weak seed beam as a function of the two photon detuning $\delta$. 
The black dashed line reports the experimental data for a heated  Rb cell, as published in ref.\cite{McCormick2007b} by the NIST group (courtesy of P. D. Lett). \red{These results are} obtained with a pump power $P$=400~mW, a pump beam waist of 650 $\mu$m ($\Omega/2\pi=330$~MHz), a pump detuning $\Delta/2\pi=700$~MHz, and $L=12.5$~mm.
The red solid line corresponds to the theoretical predictions of our model using those parameters and an atomic density  $\mathcal{N}= 4\times10^{12}$~cm$^{-3}$ in agreement with that estimated from  the measured rubidium cell temperature. The gray shaded area corresponds to the spectrum reported in Fig.~\ref{gain}a (please note the reversed scale).
The absorption dip and the Fano-like profile on the right part of the figure, absent in the simulation of Fig.~\ref{gain}, are originated by the seed acting on the $\ket{1} \to \ket{4}$ transition.
}
 \label{superpo}
\end{figure}

\indent The presented numerical results  were obtained assuming a stationary regime (no transient phenomena associated to transit-time) and in the absence of inhomogeneous Doppler broadening.
The extension of the model in order to include these effects is beyond the scope of the present work.
However, let us present the result of the calculation of the  $G_a$ seed gain, using our cold atom model, assuming the set of atom/laser parameters explored by McCormick {\it et al}~\cite{McCormick2007}.
For the atomic density, we assume $\mathcal{N} \simeq 4.10^{12}$~cm$^{-3}$ falling into the range explored by \red{the experiments of refs.~\cite{McCormick2007,McCormick2007b,McCormick2008,Pooser2008,Boyer2008,Marino2009,Glorieux2010}.}
We estimate the decoherence rate $\gamma/2\pi  =500$ kHz  from the mean transit time through the pump beam. \\
\indent Fig. \ref{superpo} shows our numerical result for  $G_a$ spectrum and compares them to the experimental observations kindly provided by P.~D.~Lett.
Notice that the gain spectrum contains twice the absorption and Fano-like feature of Fig. 2, because  the seed and conjugate beams exchange their roles in a large scan of the seed laser frequency.
 Even if the cold-atom model cannot reproduce exactly the  data obtained in a heated cell where the Doppler effect plays a major role,  the main features of the experimental transmission profile are  well reproduced. 
 In addition the gain peaks occur at the predicted two-photon detuning; the coherent absorption dips are present and finally the asymmetric profile of the left peak is also accounted for.
The gain peak values of the hot vapor experiment are in very good agreement with  those predicted by the cold-atom simulation.

\section{Conclusion}

We explored analytically and numerically the FWM generation process for an atomic double-$\Lambda$ system.
The Heisenberg-Langevin formalism allowed us to describe the classical and quantum properties of seed and conjugate beams beyond the linear noiseless amplifier approximation.
 A continuous variable approach gave us access to relative-intensity noise spectra (that can be directly compared to experiments).
 In the very wide parameter space of this system, we explored a range of large pump Rabi frequency and detuning, recently addressed by hot Rb vapors experiments.
 However, Doppler broadening was not taken into account in our analysis and the optical depth was assumed compatible with a cold-atom sample.
 In this previously unexplored regime, we predict the generation of quantum-correlated beams with a relative-intensity noise spectrum well below the standard quantum limit, down to -6 dB.
 Moreover, the calculations predict entanglement between seed and conjugate beams characterized by a inseparability down to $\mathcal{I}=0.25$.
 We verified that in the explored regime the Langevin forces cannot be neglected and reduce the performances of the system.
We examined the role played by the ground-state (hyperfine) coherence in the generation of quantum correlations:  the smooth behavior in the low $\gamma$ region let us predict that the system is robust against mild decoherence mechanisms, as magnetic field inhomogeneities.
 
Our model   investigating the double $\Lambda$ scheme  is quite general.
It may be applied  to describe a wide range of systems having the same level structure, for instance in solid state physics.

\section{Acknowledgments}
We thank C. Fabre for enlightening discussion on the symetrization problems related to calculations of Langevin forces.
We are grateful to P.~D.~Lett for discussions and for providing us unpublished results that we compare to theoretical predictions.
Finally, we thank P. Milman, O. Mishina, N. Sangouard and C. Simon for fruitful discussions at the early stage of this work.

\section*{APPENDIX}
\appendix
\section{\label{app:2}Steady state}
In zeroth-order perturbation expansion, in which $\hat a$ and $\hat b^\dag$ go to zero, the Heisenberg-Langevin equations for $\hat\sigma_{11},\hat\sigma_{22},\hat\sigma_{33},\hat\sigma_{31},\hat\sigma_{13},\hat\sigma_{42},\hat\sigma_{24}$  atomic operators are decoupled.
The mean values of these operators are required for the next order solution.
We assume the pump beam to propagate without depletion, as we verified numerically.
Then the subset of equations for the mean value variables $\mean{\hat\sigma_{11}},\mean{\hat\sigma_{22}},\mean{\hat\sigma_{33}},\mean{\hat\sigma_{31}},\mean{\hat\sigma_{13}},\mean{\hat\sigma_{42}},\mean{\hat\sigma_{24}}$ to be solved at the steady state is written in matricial form as following:
\begin{equation}\label{evol_stat}
\left(i[\mathbf{1}]\frac{\partial}{\partial{t}}+[\mean{M_0}]\right)|\Sigma_0]=|S_0]
\end{equation}
with \begin{widetext}
 \begin{equation}
[M_0]=\left(
\begin{array}{ccccccc}
 i \frac{\Gamma }{2} & i \frac{\Gamma }{2} & 0 & -\frac{\Omega }{2} & \frac{\Omega }{2} & 0 & 0 \\
 i \frac{\Gamma }{2} & i \frac{\Gamma }{2} & 0 & 0 & 0 & -\frac{\Omega }{2} & \frac{\Omega }{2} \\
 0 & 0 & i \Gamma  & \frac{\Omega }{2} & -\frac{\Omega }{2} & 0 & 0 \\
 -\frac{\Omega }{2} & 0 & \frac{\Omega }{2} & -\Delta +i \frac{\Gamma }{2} & 0 & 0 & 0 \\
 \frac{\Omega }{2} & 0 & -\frac{\Omega }{2} & 0 & \Delta +i \frac{\Gamma }{2} & 0 & 0 \\
 -\frac{\Omega }{2} & -\Omega  & -\frac{\Omega }{2} & 0 & 0 & -\Delta -\omega_0+i \frac{\Gamma }{2} & 0 \\
 \frac{\Omega }{2} & \Omega  & \frac{\Omega }{2} & 0 & 0 & 0 & \Delta +\omega_0+i \frac{\Gamma }{2}
\end{array}
\right)\\
\end{equation}
\end{widetext}
\begin{equation}
|\Sigma_0]=\left(
\begin{array}{c}
\sigma_{11}\\
\sigma_{22}\\
\sigma_{33}\\
\sigma_{31}\\
\sigma_{13}\\
\sigma_{42}\\
\sigma_{24}
\end{array}
\right),\ 
|S_0]=\frac{1}{2}\left(
\begin{array}{c}
 i \Gamma  \\
 i \Gamma  \\
 0 \\
 0 \\
 0 \\
 -\Omega  \\
 \Omega 
\end{array}
\right).
\end{equation}
The steady state solution of Eq.~\eqref{evol_stat} is 
\begin{equation}
|\langle\Sigma_0\rangle]=[M_0]^{-1}|S_0].
\end{equation}

\section{\label{app:3}Atomic Heisenberg-Langevin equations}
The first order solution for the four coherences, $\hat\sigma_{23},\hat\sigma_{41},\hat\sigma_{43},\hat\sigma_{21}$ is determined by the following  matricial equation
\begin{equation}\label{evol_cohe}
\left(i[\mathbf{1}]\frac{\partial}{\partial{t}}+[M_1]\right)|\Sigma_1]=|S_1]|\hat{A}]+i|F_1]
\end{equation}
with\begin{widetext}
\begin{eqnarray}
[M_1]=\left[\begin{array}{cccc}
i \Gamma/2+(\Delta -\delta)  & 0 & -\Omega/2 & \Omega/2 \\
 0 & i \Gamma/2-(\Delta +\delta +\omega_0) & \Omega/2 & -\Omega/2 \\
 -\Omega/2 & \Omega/2 &i\Gamma -(\delta +\omega_0) & 0 \\
 \Omega/2 & -\Omega/2 & 0 &  i \gamma -\delta 
\end{array}
\right]
\end{eqnarray}
\end{widetext}
\begin{eqnarray}
&&\nonumber|\Sigma_1]=\left[
\begin{array}{c}
\sigma_{23}\\
\sigma_{41}\\
\sigma_{43}\\
\sigma_{21}
\end{array}
\right],\ 
|S_1]=g\left[
\begin{array}{cc}
\langle\sigma_{33}-\sigma_{22}\rangle&0 \\
0 &\langle\sigma_{11}-\sigma_{44}\rangle\\
 -\langle\sigma_{42}\rangle&\langle\sigma_{13}\rangle \\
 \langle\sigma_{31}\rangle &-\langle\sigma_{24}\rangle
\end{array}
\right],
\\
&&|\hat{A}]=\left[
\begin{array}{c}
\hat{a}\\
\hat{b}^\dag
\end{array}
\right],\ 
|F_1]=\left[
\begin{array}{c}
F_{23}\\
F_{41}\\
F_{43}\\
F_{21}
\end{array}
\right].
\end{eqnarray}
A similar set of equations holds for the hermitian conjugate operators.

{The Langevin atomic forces $|F]$ are characterized by their diffusion coefficients matrix $[D]$ (Eq.~\eqref{diffusion}). 
In order to comply with the symmetrical ordering condition introduced in Section \ref{sec:results}, we write the following symmetrize diffusion coefficients}
\begin{equation}
[D]=([D_1]+[D_2]),
\label{atomicdiffusion}
\end{equation}
having defined
\begin{equation}
[D_1]  2 \delta(t-t')\delta(z-z')=\langle|F_1(z,t)] [F_1^\dag(z,t')|\rangle,
\end{equation}
\begin{equation}
[D_2]  2\delta(t-t')\delta(z-z')=\langle|F_1^\dag(z,t)] [F_1(z,t')|\rangle.
 \end{equation}
\indent Langevin diffusion coefficients for operators can be calculated using the \textit{generalized Einstein relation} as in ref.~\cite{Davidovich1996}.
 The $[D_1]$ and $[D_2]$  diffusion matrices  are given by
\begin{widetext}
\begin{footnotesize}
\begin{eqnarray}
\nonumber [D_1]&=&\frac{1}{2\tau}\left[
\begin{array}{cccc}
 \Gamma  \left(\Gamma ^2+4 \Delta ^2+2 \Omega ^2+8 \Delta  \omega_0+4 \omega_0^2\right) & 0 & i \Gamma  \Omega  (\Gamma +2 i (\Delta +\omega_0)) & 0 \\
 0 & 0 & 0 & -i \gamma  \Omega  (\Gamma -2 i (\Delta +\omega_0)) \\
 -i \Gamma  \Omega  (\Gamma -2 i (\Delta +\omega_0)) & 0 & \Gamma  \Omega ^2 & 0 \\
 0 & i \gamma  \Omega  (\Gamma +2 i (\Delta +\omega_0)) & 0 & \Gamma  \Omega ^2+2 \gamma  \left(\Gamma ^2+4 \Delta ^2+\Omega ^2+8 \Delta  \omega_0+4 \omega_0^2\right)
\end{array}\right]\\
~[D_2]&=&\frac{1}{2\tau}\left[
\begin{array}{cccc}
 0 & 0 & 0 & -i \gamma  (\Gamma -2 i \Delta ) \Omega  \\
 0 & \Gamma  \left(\Gamma ^2+4 \Delta ^2+2 \Omega ^2\right) & i \Gamma  (\Gamma +2 i \Delta ) \Omega  & 0 \\
 0 & -i \Gamma  (\Gamma -2 i \Delta ) \Omega  & \Gamma  \Omega ^2 & 0 \\
 i \gamma  (\Gamma +2 i \Delta ) \Omega  & 0 & 0 & \Gamma  \Omega ^2+2 \gamma  \left(\Gamma ^2+4 \Delta ^2+\Omega ^2\right)
\end{array}
\right]
\end{eqnarray}
\end{footnotesize}
where 
 $\tau=2 \Gamma ^2+4\Omega ^2+4\omega_0^2+8 \Delta ^2+8 \Delta  \omega_0$
 \end{widetext}

The system of Eq. \eqref{evol_cohe} is solved by writing each atomic operator as the sum of its mean value and a quantum fluctuation term
\begin{equation}
|\Sigma_1]=|\langle\Sigma_1\rangle]+|\delta\Sigma_1],
\label{operators}
\end{equation}
By linearizing Eq.~\eqref{evol_cohe} we derive for the mean values
\begin{equation}
|\langle\Sigma_1\rangle]=[M_1]^{-1}|S_1]|\langle\hat{A}\rangle].
\end{equation}
and for the Fourier transformed quantum fluctuations
\begin{equation}
|\delta\Sigma_1]=([M_1]+\omega[\mathbb{I}])^{-1}|S_1]|\delta\hat{A}]+i([M_1]+\omega[\mathbb{I}])^{-1}|F_1].
\label{flucuat}
\end{equation}

\section{Quantum field Heisenberg-Langevin equations\label{app:C}}

The propagation matrix for the quantum field operators of Eq.~\eqref{eq:8} is obtained by replacing the $\hat\sigma_{23}(t,z)$ and $\hat\sigma_{41}(t,z)$ solutions of Eq.\eqref{operators} into the differential Eqs.~\eqref{eq:propa1} - \eqref{eq:propa2} and solving these equations.  By defining
\begin{equation}
\begin{bmatrix}A(\omega) & B(\omega) \\C(\omega) & D(\omega)\end{bmatrix}=e^{[M(\omega)]L},
\end{equation}
where $[M(\omega)]$ is a 2x2 matrix given by
\begin{equation}
 [M(\omega)]=-i \frac{g^2 N}{c} [T]([M_1]+\omega[\mathbb{I}])^{-1}|S_1]
 \end{equation}
 with  $[T]=\begin{bmatrix}
-1 & 0 & 0 & 0 \\ 
0 & 1 & 0 & 0
\end{bmatrix}$.\\
The Langevin force terms of Eq.~\eqref{eq:8} are given by
\begin{equation}
\begin{bmatrix}\hat{F}_{a,out}(\omega)\\\hat{F}_{b^\dag,out}(\omega)\end{bmatrix} =L\int^1_0 e^{[M(\omega)]Lz}[M_F(\omega)]|F_1]dz
\end{equation}
where
\begin{equation}
[M_F(\omega)]=-\frac{g N}{c} [T]([M_1]+\omega[\mathbf{1}])^{-1}.
\end{equation}
Let us point out that the conjugate operators for the quantum fields are given by
 \begin{equation}\label{adjoint}
\left[
 \begin{array}{c}
 \delta \hat a^\dag_{out}(\omega)  \\ 
 \delta \hat b_{out}(\omega) 
 \end{array} \right]=e^{[M^*(-\omega)]L} \left(\begin{bmatrix} \delta\hat{a}_{in}(\omega) \\ \delta\hat{b}_{in}^\dag(\omega)\end{bmatrix}+
\begin{bmatrix}\hat{F}_{a^\dag,out}(\omega)\\\hat{F}_{b,out}(\omega)\end{bmatrix}\right),
 \end{equation}
 			with 	 
\begin{equation}\label{def_Fdag}
 		\left[
 \begin{array}{c}
 F_{a^\dag,out}(\omega)  \\ 
 F_{b,out}(\omega) 
 \end{array} \right]=L\int_0^1  e^{-[M^*(-\omega)]Lz} [M_F^*(-\omega)]|F_1^\dag]dz,
\end{equation} 
where we have introduced $|F_1^\dag]=\left[
\begin{array}{c}
F_{32}\\
F_{14}\\
F_{34}\\
F_{12}
 \end{array} \right].
$

\section{\label{appendix_langevin} Langevin forces contribution to noise spectra}
We will present this calculation for the term $\langle \hat F_{a}(\omega)\hat F_{a^\dag}(\omega ')\rangle$ term, as the extension to  the remaining ones is straightforward. 
Substituting the expression of Eq. \eqref{def_Fdag} this term is given by
\begin{eqnarray}
\nonumber\langle \hat F_{a}(\omega)\hat F_{a^\dag}(\omega ')\rangle=\langle[1\ 0|\int_0^1  e^{-[M(\omega)]Lz} [M_F(\omega)]|F_1(z,\omega)]dz\\
 \times  \int_0^1 [F_1^\dag(z',\omega')|\ ^t[M_{F}^*(-\omega')]e^{-^t[M^*(-\omega')]Lz'}dz' |1\ 0]\rangle\hspace{0.3cm}
\end{eqnarray}
The Langevin forces are $\delta$-correlated in $z$ so integration over $z'$ gives :
\begin{eqnarray}
\nonumber\langle \hat F_{a}(\omega)\hat F_{a^\dag}(\omega ')\rangle =[1\ 0|\langle\int_0^1  e^{-[M(\omega)]Lz}[M_{F}(\omega)]|F_1(z,\omega)]\\
\times  [F_1^\dag(Lz,\omega')|\ ^t[M_{F}^*(-\omega')]e^{-^t[M^*(-\omega')]Lz}dz\rangle|1\ 0].\hspace{0.5cm}
\end{eqnarray}

By defining the $D_{aa^\dag}(\omega)$ coefficient as
\begin{equation}
\langle \hat F_{a}(\omega)\hat F_{a^\dag}(\omega ')\rangle =D_{aa^\dag}(\omega) 2\pi\ \delta(\omega+\omega')
\end{equation}
we obtain 
\begin{eqnarray}
\nonumber D_{aa^\dag}(\omega)=L^2 [1\ 0|\int_0^1  e^{-[M(\omega)]Lz}[M_{F}(\omega)][D]\\
\times ^t[M_{F}^*(-\omega')]e^{-^t[M^*(-\omega')]Lz}dz\rangle|1\ 0],
\end{eqnarray}
within the $[D]$ diffusion matrix given by Eq.~\eqref{atomicdiffusion}. In a similar way it is possible to derive  $ D_{a^\dag a}(\omega)$,  $ D_{bb^\dag}(\omega)$ and  $ D_{b^\dag b}(\omega)$, as detailed in \cite{GlorieuxPhD}.

\end{document}